\documentclass[aps,pra,reprint,showpacs,superscriptaddress,notitlepage,twocolumn]{revtex4-1}%
\usepackage{amsfonts}
\usepackage{mathrsfs}
\usepackage{amsmath}
\usepackage{amssymb}
\usepackage{graphicx}
\usepackage{color}%
\usepackage{mathtools}
\usepackage{booktabs}
\usepackage[colorlinks,linkcolor=blue,citecolor=blue,hyperindex,bookmarks=false,pdfstartview=FitH]{hyperref}

\providecommand{\U}[1]{\protect\rule{.1in}{.1in}}

\newcommand{\figpanel}[2]{\hyperref[#1]{\ref*{#1}(#2)}}

\begin{document}
\title{A giant atom with modulated transition frequency}

\author{Lei Du}
\affiliation{Center for Theoretical Physics and School of Science, Hainan University, Haikou 570228, China}
\affiliation{Beijing Computational Science Research Center, Beijing 100193, China}
\author{Yan Zhang}
\email{zhangy345@nenu.edu.cn}
\affiliation{Center for Quantum Sciences and School of Physics, Northeast Normal University, Changchun 130024, China}
\author{Yong Li}
\email{yongli@hainanu.edu.cn}
\affiliation{Center for Theoretical Physics and School of Science, Hainan University, Haikou 570228, China}
\affiliation{Synergetic Innovation Center for Quantum Effects and Applications, Hunan Normal University, Changsha 410081, China}


\begin{abstract} 
Giant atoms are known for the frequency-dependent spontaneous emission and associated interference effects. In this paper, we study the spontaneous emission dynamics of a two-level giant atom with dynamically modulated transition frequency. It is shown that the retarded feedback effect of the giant-atom system is greatly modified by a dynamical phase arising from the frequency modulation and the retardation effect itself. Interestingly, such a modification can in turn suppress the retarded feedback such that the giant atom behaves like a small one. By introducing an additional phase difference between the two atom-waveguide coupling paths, we also demonstrate the possibility of realizing chiral and tunable temporal profiles of the output fields. The results in this paper have potential applications in quantum information processing and quantum network engineering.
  
\end{abstract}

\maketitle

\section{Introduction}

Spontaneous emission is a basic and important process that arises from the interaction between an excited quantum system and the surrounding environment~\cite{AgarwalSE,OpenSys,QuantNoise}. This process is typically irreversible and thus plays a negative role in quantum information processing, e.g., leading to the so-called quantum decoherence effect. On the other hand, controlling the spontaneous emission of an open quantum system has great significance for many applications, such as quantum switch engineering~\cite{JTShen2005,CPSun1,CPSun2,Baranger2013,AGTrouter,OPainter}, high-frequency coherent light generation~\cite{hflg1,hflg2}, clock frequency estimation~\cite{clock}, and chiral quantum optics~\cite{Lodahlchiral,Zollerchiral}. As a result, the control of spontaneous emission has attracted a plethora of research interest, with common strategies including designing the density of states of the reservoir~\cite{cavity1,cavity2,cavity3,dos1,dos2,dos3,dos4}, changing the system-reservoir couplings~\cite{chancp1,chancp2,chancp3}, and using strong pulse sequences~\cite{pulse1,pulse2,pulse3} or low-frequency coherent fields~\cite{cofield1,cofield2}, to name a few.

Recent advances in waveguide quantum electrodynamics demonstrate that a single quantum emitter can be coupled to the waveguide field at multiple separate points. In this case, the spontaneous emission of the emitter can be significantly modified, depending on both the transition frequency of the emitter and the spacing distances between different coupling points~\cite{LambAFK}. Such a new quantum optical paradigm is referred to as ``giant atoms''~\cite{fiveyear}, whose size can be much larger than the wavelength of the field. In this case, the propagation phases of the field between different coupling points should be considered, since they play a vital role in determining the spontaneous emission of the emitter. This resembles a small atom in front of a mirror~\cite{chancp2,SemiDH,SemiShen,FCretard1,FCretard2,FCretard3}, which can be mapped to a giant one with two identical coupling paths. However, giant atoms typically allow for more intriguing interference effects and advanced scattering behaviors due to their richer geometries~\cite{fiveyear}. To date, giant atoms have witnessed a series of intriguing quantum optical phenomena, such as decoherence-free atomic interactions~\cite{NoriGA,braided,FCdeco}, unconventional bound states~\cite{oscillate,BICGA,WXchiral1,ZhaoWbound,VegaPRA}, chiral quantum optics~\cite{AFKchiral,WXchiral1,WXchiral2}, synthetic dimension manipulation~\cite{DLsyn}, and phase-dependent single-photon scatterings~\cite{DLlambda,DLprr,JiaGA1,JiaGA2,ZhaoFOP,YinGA,CYTcp}, and photon storage~\cite{YuanGA}.

In this paper, we consider a two-level giant atom with modulated transition frequency. If the system is operated in the non-Markovian regime, where the propagation time of the field (e.g., photons) between the two atom-waveguide coupling points cannot be neglected compared with the relaxation time of the atom, the frequency modulation imprints a time-dependent modification on the retarded feedback and thereby changes the spontaneous emission dynamics of the system. The modification effect depends on both the concrete form of the frequency modulation and the propagation time between the two coupling points. Such a modification tends to disappear if the system enters the Markovian regime with negligible propagation time. As will be shown below, the combination of the giant-atom interference effect (i.e., frequency-dependent Lamb shift and relaxation rate of the giant atom due to the interference of the multiple atom-waveguide coupling paths~\cite{LambAFK}) and the frequency modulation not only enables richer non-Markovian dynamics, but also shows the possibility of engineering chiral single-photon source with tunable temporal profiles.    

\section{Model and method}\label{sec2}

We consider in this paper a two-level giant atom whose transition frequency $\omega(t)$ is dynamically modulated in the vicinity of a constant value $\omega_{0}$ (i.e., $|\omega(t)-\omega_{0}|\ll\omega_{0}$, which justifies the linearized dispersion relation of the waveguide employed below). Experimentally, two-level systems with modulated transition frequencies can be implemented via, e.g., driven superconducting qubits~\cite{expsq1,expsq2} or quantum dots~\cite{expqd}. Assuming that the atom is coupled to the one-dimensional waveguide at two points $x=0$ and $x=d$, the Hamiltonian of the system can be written as ($\hbar=1$ hereafter)
\begin{equation}
\begin{split}
H(t)&=\omega(t)\sigma_{+}\sigma_{-}+\int_{-\infty}^{+\infty}dk\omega_{k}a_{k}^{\dag}a_{k}\\
&\quad\,+\int_{-\infty}^{+\infty}dk\left[g\left(1+e^{i\varphi}e^{ikd}\right)\sigma_{+}a_{k}+\text{H.c.}\right],
\end{split}
\label{eq1}
\end{equation}
where $\sigma_{+}$ ($\sigma_{-}$) is the raising (lowering) operator of the two-level atom; $a_{k}^{\dag}$ ($a_{k}$) is the creation (annihilation) operator of the waveguide mode with frequency $\omega_{k}$ and wave vector $k$; $g$ is the coupling coefficient between the atom and the waveguide, which is assumed to be $k$ independent and identical at the two coupling points. Note that in Eq.~(\ref{eq1}), we have used the rotating-wave approximation and the spectrum of the waveguide field in the continuum limit. Moreover, we have introduced an additional phase difference $\varphi$ between the two atom-waveguide coupling paths at $x=0$ and $x=d$, which can be achieved experimentally via a number of artificial methods, such as dynamically modulating the coupler between the atom and the waveguide~\cite{WXchiral1,WXchiral2,phasediff1,phasediff2,phasediff3} and introducing an optical path difference for opposite directions of photon hopping~\cite{phasediff4,phasediff5}. Such a phase difference mimics a synthetic gauge field so that it imprints a momentum kick on the emitted photons. In this case, the atom-waveguide interaction and thereby the spontaneous emission of the atom becomes chiral~\cite{ZollerAB}. 

It is clear that the total excitation number of the system, which is defined by the operator $N=\int dka_{k}^{\dag}a_{k}+\sigma_{+}\sigma_{-}$, is conserved due to $[N,\,H(t)]=0$. Therefore, if the system is initialized in a single-excitation state, the state of the system at time $t>0$ can be written as 
\begin{equation}
\begin{split}
|\psi(t)\rangle&=\int_{-\infty}^{+\infty}dk c_{k}(t)a_{k}^{\dag}e^{-i\omega_{k}t}|G\rangle\\
&\quad\,+c_{e}(t)\sigma_{+}e^{-i\int_{0}^{t}dt''\omega(t'')}|G\rangle,
\end{split}
\label{eq2}
\end{equation}
where $c_{k}(t)$ [$c_{e}(t)$] is the probability amplitude of creating a photon with wave vector $k$ in the waveguide (of the atom in the excited state); $|G\rangle$ denotes that the atom is in the ground state and there is no photon in the waveguide. In this paper, we focus on the spontaneous emission dynamics of the modulated giant atom, thus the initial state is always assumed to be $|\psi(0)\rangle=\sigma_{+}|G\rangle$, i.e., the atom is in the excited state and the waveguide is empty at the initial time. By solving the Schr\"{o}dinger equation, one obtains
\begin{eqnarray}
\dot{c}_{e}(t)&=&-i\int_{-\infty}^{+\infty}dk g\left(1+e^{i\varphi}e^{ikd}\right)c_{k}(t)\nonumber\\
&&\times e^{-i\omega_{k}t}e^{i\int_{0}^{t}dt''\omega(t'')},\label{eq3}\\
\dot{c}_{k}(t)&=&-ig\left(1+e^{-i\varphi}e^{-ikd}\right)c_{e}(t)\nonumber\\
&&\times e^{i\omega_{k}t}e^{-i\int_{0}^{t}dt''\omega(t'')}.\label{eq4}
\end{eqnarray} 
Substituting the formal solution of Eq.~(\ref{eq4}), i.e.,
\begin{equation}
\begin{split}
c_{k}(t)&=-i\int_{0}^{t}dt' g\left(1+e^{-i\varphi}e^{-ikd}\right)\\
&\quad\,\times c_{e}(t')e^{i\omega_{k}t'}e^{-i\int_{0}^{t'}dt''\omega(t'')},
\end{split}
\label{eq5}
\end{equation}
into Eq.~(\ref{eq3}), one arrives at
\begin{equation}
\begin{split}
\dot{c}_{e}(t)&=-2\int_{0}^{t}dt'\int_{-\infty}^{+\infty}dk g^{2}[1+\cos{(\varphi+kd)}]\\
&\quad\,\times c_{e}(t')e^{-i\omega_{k}(t-t')}e^{i\int_{t'}^{t}dt''\omega(t'')}.
\end{split}
\label{eq6}
\end{equation}
Note that we have taken $c_{k}(0)=0$ in Eq.~(\ref{eq5}) due to the initial state $|\psi(0)\rangle=\sigma^{+}|G\rangle$ and exchanged the integration order in Eq.~(\ref{eq6}) as usual~\cite{QuantNoise}. By assuming $\omega_{k}\approx\omega_{0}+\nu_{k}=\omega_{0}+(k-k_{0})v_{g}$~\cite{JTShen2005,JTShen2009}, where $v_{g}$ ($k_{0}$) is the group velocity (wave vector) of the field at frequency $\omega_{0}$, and changing the integration variable as $\int_{-\infty}^{+\infty}f(k)dk\rightarrow\int_{0}^{+\infty}[f(k)+f(-k)]d\omega_{k}/v_{g}\rightarrow\int_{-\infty}^{+\infty}[f(k)+f(-k)]d\nu_{k}/v_{g}$, Eq.~(\ref{eq6}) becomes 
\begin{widetext}
\begin{equation}
\begin{split}
\dot{c}_{e}(t)&=-\frac{2g^{2}}{v_{g}}\int_{0}^{t}dt'\int_{-\infty}^{+\infty}d\nu_{k}\left[2+\cos{\varphi}\left(e^{ik_{0}d}e^{i\frac{\nu_{k}}{v_{g}}d}+e^{-ik_{0}d}e^{-i\frac{\nu_{k}}{v_{g}}d}\right)\right]c_{e}(t')e^{-i(\omega_{0}+\nu_{k})(t-t')}e^{i\int_{t'}^{t}dt''\omega(t'')}\\
&=-\Gamma\int_{0}^{t}dt' c_{e}(t')\left\{2\delta(t-t')+\cos{\varphi}\left[e^{ik_{0}d}\delta(t-t'-\tau)+e^{-ik_{0}d}\delta(t-t'+\tau)\right]\right\}e^{-i\omega_{0}(t-t')}e^{i\int_{t'}^{t}dt''\omega(t'')}\\
&=\underbrace{-\Gamma c_{e}(t)}_{\text{instantaneous\,\,decay}}\underbrace{-\Gamma\cos{\varphi}e^{i\phi(t,\tau)}c_{e}(t-\tau)\Theta(t-\tau)}_{\text{retarded\,\,feedback}}
\end{split}
\label{eq7}
\end{equation}
\end{widetext}
with 
\begin{equation}
\phi(t,\tau)=\phi_{0}-\omega_{0}\tau+\int_{t-\tau}^{t}dt'\omega(t'),
\label{eq8}
\end{equation}
where $\Gamma=4\pi g^{2}/v_{g}$ is the radiative decay rate of the atom; $\phi_{0}=k_{0}d$ describes a static phase accumulation that exists even without the modulation~\cite{LonghiGA,GLZ2017}; $\tau=d/v_{g}$ is the time delay (propagation time) of photons traveling between the two coupling points; $\Theta(x)$ is the Heaviside step function. Note that we have extended the integration limit of $\nu_{k}$ to $(-\infty,+\infty)$. This is justified since the intensity of the atomic power spectrum is concentrated around the bare transition frequency $\omega_{0}$ and thus the extended part makes negligible contribution to the integral~\cite{JTShen2009}. Moreover, the terms containing $\sin{\varphi}$ disappear in the first line of Eq.~(\ref{eq7}) due to $\sin{(-kd)}=-\sin{(kd)}$. Clearly, Eq.~(\ref{eq7}) describes a non-Markovian dynamics with a retarded coherent feedback. In contrast to the common situation where the transition frequency of the giant atom is constant and there is no additional phase difference between the two coupling paths, the present model has two interesting hallmarks: (i) the retarded feedback term contains a dynamical phase $\phi(t,\tau)$ that is determined by the concrete form of $\omega(t)$ as well as the value of $\tau$; (ii) the amplitude of the feedback term is further modified by the additional phase difference $\varphi$ in terms of a cosine function. Note that $\phi(t,\tau)$ becomes \emph{trivial} if $\tau$ is exactly zero since in this case the model reduces to a small-atom system with $d=0$. 

In the case of $\omega(t)\equiv\omega_{0}$ and $\varphi=0$, Eq.~(\ref{eq7}) becomes
\begin{equation}
\dot{c}_{e}(t)=-\Gamma c_{e}(t)-\Gamma c_{e}(t-\tau)e^{i\phi_{0}}\Theta(t-\tau),
\label{eq9}
\end{equation}
which recovers the dynamic equation of a common giant atom that has been well studied previously~\cite{LonghiGA,GLZ2017}. In this case, the retardation effect simply postpones the onset of the giant-atom interference effect determined by the value of $\phi_{0}$. For example, the atom exhibits partial decay (i.e., exponential decay at the initial stage and inhibited decay for $t>\tau$) if $\phi_{0}=(2m+1)\pi$ ($m$ is an arbitrary integer). This can be seen by solving the Laplace transformation of Eq.~(\ref{eq9}) and using the final value theorem, which yields $c_{e}(t\rightarrow+\infty)=1/(1+\Gamma\tau)$ in this case. In particular, when $\tau=0$, the spontaneous emission of the giant atom is completely inhibited, implying that the atom is in a dark state in this case. This phenomenon is however different from the subradiance behaviors of multiple small atoms~\cite{sub1,sub2}. While the latter is a collective effect and occurs with specific initial conditions, the inhibited spontaneous emission of a single giant atom is only related to the phase accumulations between different coupling points. Such a dark state cannot be obtained if $\text{mod}(\varphi,\pi)\neq0$, since the retarded feedback (whose amplitude is proportional to $\cos{\varphi}$) cannot exactly cancel the instantaneous decay of the atom in this case.   

\section{Controllable spontaneous emission}\label{sec3}   

In this section, we consider a simple cosine-type modulation 
\begin{equation}
\omega(t)=\omega_{0}+\alpha\cos{(\Omega t+\theta)} 
\label{eq10}
\end{equation}
around the background frequency $\omega_{0}$, where $\alpha$, $\Omega$, and $\theta$ are the amplitude, frequency, and initial phase of the modulation, respectively. In this case, the dynamical phase $\phi(t,\tau)$ can be written as
\begin{equation}
\phi(t,\tau)=\phi_{0}+\chi[\sin{(\Omega t+\theta)}-\sin{(\Omega t-\Omega\tau+\theta})]
\label{eq11}
\end{equation} 
with $\chi=\alpha/\Omega$ the modulation depth~\cite{Keitel2014}. It is clear from Eq.~(\ref{eq11}) that the spontaneous emission dynamics of the giant atom can be controlled by tuning $\Omega$, $\chi$, and $\theta$. To study the non-Markovian retardation effect, we consider a finite time delay ($\Gamma\tau=0.2$) that is nonnegligible compared with the relaxation time of the atom. Moreover, we assume $\varphi=0$ tentatively in order to highlight the effect of the frequency modulation. We will discuss the influence of the additional phase difference $\varphi$ on the spontaneous emission dynamics of the atom at the end of this section and on the output fields of the model in the next section.

\begin{figure}[ptb]
\centering
\includegraphics[width=8.5 cm]{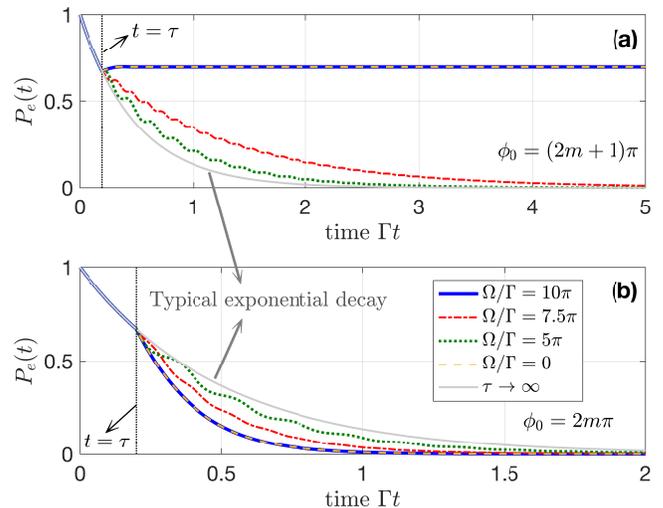}
\caption{Dynamic evolutions of atomic population probability $P_{e}(t)$ with (a) $\phi_{0}=(2m+1)\pi$ and (b) $\phi_{0}=2m\pi$. Panels (a) and (b) share the same legend. The vertical dotted lines correspond to the moment $t=\tau$ that the atom feels the retarded feedback. Other parameters are $\chi=1$, $\theta=0$, $\varphi=0$, and $\tau\Gamma=0.2$ (except for indicated).}\label{fig1}
\end{figure}

We first focus on the influence of the modulation frequency and plot in Fig.~\ref{fig1} the dynamic evolutions of the atomic population probability $P_{e}(t)=|c_{e}(t)|^{2}$ for different values of $\Omega$. For $\phi_{0}=(2m+1)\pi$, as shown in Fig.~\figpanel{fig1}{a}, the cosine-type modulation markedly modifies the dynamics of the atom. When $\Omega\tau=2n\pi$ ($n$ is another arbitrary integer that is in general unequal to $m$), the spontaneous emission of the atom is inhibited after $t=\tau$, just as it would be in the absence of the modulation (see the coincident blue solid and yellow dashed lines). As $\Omega\tau$ approaches $(2n-1)\pi$, the decay of the atom tends to be exponential-like but with a slight oscillation. This can be well understood from Eq.~(\ref{eq11}): when $\Omega\tau=2n\pi$, $\phi(t,\tau)\equiv\phi_{0}$ becomes time independent as if there is no modulation; for other cases of $\Omega\tau\neq2n\pi$, $\phi(t,\tau)$ changes periodically in time due to the cosine-type modulation. Moreover, as shown in Fig.~\figpanel{fig1}{b}, one can tune the dynamics of the atom between the superradiant-like form (arising from the constructive interference between the two atom-waveguide coupling paths; see the blue solid and yellow dashed lines) and the typical exponential form (the giant-atom interference effect never takes place for $\tau\rightarrow\infty$; see the gray solid line) in the case of $\phi_{0}=2m\pi$. Note that $\tau=0$ ($d=0$) corresponds to a small atom as well, but in this case the atomic decay rate is twofold (i.e., $4\Gamma$) due to the constructive interference between the two overlapped coupling points.  

A recent work~\cite{DLprr2} has shown that by modulating the atom-waveguide coupling strengths, it is also possible to control the spontaneous emission dynamics of a giant atom. Indeed, both the frequency and coupling modulation schemes require the system to be in the non-Markovian regime and the atomic dynamics exhibit similar behaviors when considering cosine-type modulations in these two cases. From this point of view, frequency modulations provide an alternative wisdom for controlling the spontaneous emission dynamics of a giant atom. For some platforms, however, modulating the atomic frequency is much easier than altering the interaction between the atom and the waveguide (such as natural atoms coupled to the evanescent fields of optical fibers). Moreover, it is challenging to introduce modulations precisely to all the coupling paths of the giant atom, especially when the number of the coupling points is very large. Nevertheless, as will be shown below, there is a limitation on the effect of the present scheme. 

\begin{figure}[ptb]
\centering
\includegraphics[width=8.5 cm]{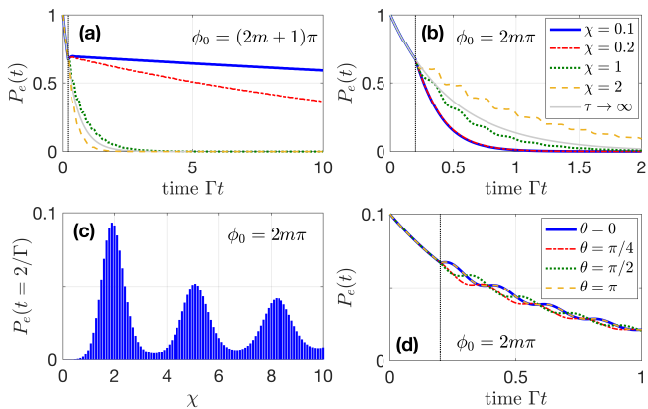}
\caption{\textbf{(a, b)} Dynamic evolutions of atomic population probability $P_{e}(t)$ with different values of $\chi$ and $\phi_{0}$. Panels (a) and (b) share the same legend. \textbf{(c)} $P_{e}(t=2/\Gamma)$ as a function of $\chi$ for $\phi_{0}=2m\pi$. \textbf{(d)} Dynamic evolutions of atomic population probability $P_{e}(t)$ with different values of $\theta$ and $\phi_{0}=2m\pi$. We assume $\theta=0$ in panels (a)-(c) and $\chi=1$ in panel (d). The vertical dotted lines in (a), (b), and (d) correspond to $t=\tau$ as those in Fig.~\ref{fig1}. Other parameters are $\tau\Gamma=0.2$, $\Omega/\Gamma=5\pi$, and $\varphi=0$.}\label{fig2}
\end{figure}

It can be seen from Eq.~(\ref{eq11}) that the modulation effect is strongly influenced by the modulation depth $\chi$. If $\chi$ is very small, the modulation effect is limited because the dynamical part of $\phi(t,\tau)$ changes within the finite range $[-2\chi,\,2\chi]$. This is also why the spontaneous emission of the atom cannot be further boosted (suppressed) in Fig.~\figpanel{fig1}{a} [Fig.~\figpanel{fig1}{b}]. In view of this, we examine in Figs.~\figpanel{fig2}{a} and \figpanel{fig2}{b} the evolutions of $P_{e}(t)$ for $\Omega\tau=(2n+1)\pi$ and different values of $\chi$. When $\phi_{0}=(2m+1)\pi$, as shown in Fig.~\figpanel{fig2}{a}, the atom exhibits a nearly linear decay for small $\chi$ (see, e.g., the blue solid and red dot-dashed lines) and an exponential-like decay for large $\chi$ (see, e.g., the green dotted and yellow dashed lines). This suggests a way to engineer richer spontaneous emission dynamics for quantum emitters. Similarly, when $\phi_{0}=2m\pi$ as shown in Fig.~\figpanel{fig2}{b}, the atomic decay can be further suppressed upon increasing $\chi$ properly. In Figs.~\figpanel{fig2}{a} and \figpanel{fig2}{b}, we have concentrated on the situation of $\chi\in[0,2]$, in which the modulation effect shows a monotonic behavior with the increase of $\chi$. However, it is no longer the case if we further increase the value of $\chi$. As shown in Fig.~\figpanel{fig2}{c}, the atomic population $P_{e}(t=2/\Gamma)$ exhibits a damped oscillation as $\chi$ increases and the maximum is found around $\chi=2$. Such a non-monotonic behavior can be understood from the Jacobi-Anger extension of the dynamical phase factor, which will be discussed in detail below. Moreover, Fig.~\figpanel{fig2}{d} shows that the spontaneous emission of the atom is quite insensitive to the modulation phase $\theta$. Tuning $\theta$ only leads to a slight phase shift for the oscillating evolution curve.   

Before proceeding, we would like to point out that the modulation effects shown above tend to disappear as the time delay $\tau$ decreases gradually. This can be seen again from Eq.~(\ref{eq11}): $\phi(t,\tau)\approx\phi_{0}$ if $\tau$ is much smaller than the other timescales. In view of this, the controllable spontaneous emission here is closely related to the non-Markovian retardation effect arising from the giant-atom structure. Moreover, our scheme is quite different from that in Ref.~\cite{Janowicz}, where a structured reservoir with a narrow band is required in order to suppress the spontaneous emission of a modulated small atom. In our scheme the spectrum of the waveguide modes can be very broad and flat. 


\begin{figure}[ptb]
\centering
\includegraphics[width=8.5 cm]{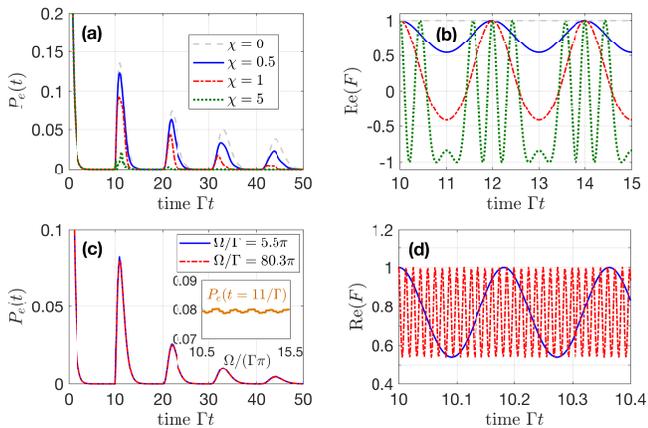}
\caption{\textbf{(a, c)} Dynamic evolutions of atomic population probability $P_{e}(t)$ with (a) different values of modulation depth $\chi$ and (c) different values of modulation frequency $\Omega$. The inset in panel (c) depicts $P_{e}(t=11/\Gamma)$ versus $\Omega$ for fixed $\chi$. \textbf{(b, d)} Dynamic evolutions of the real part of the dynamical phase factor $\text{Re}(F)$ with (b) different values of modulation depth $\chi$ and (d) different values of modulation frequency $\Omega$. We assume $\Omega/\Gamma=0.5\pi$ in panels (a) and (b) and $\chi=0.5$ in panels (c) and (d). Panels (a) and (b) [(c) and (d)] share the same legend. Other parameters are $\phi_{0}=2m\pi$, $\tau\Gamma=10$, $\theta=0$, and $\varphi=0$.}\label{fig3}
\end{figure} 

It has been shown that a common giant atom described by Eq.~(\ref{eq9}) enables periodic population revivals in the deep non-Markovian regime (i.e., $\Gamma\tau\gg1$), with adjacent revivals equally spaced by $\tau$~\cite{GLZ2017,nonexp}. This non-Markovian effect, however, can be markedly suppressed by the cosine-type frequency modulation. In Fig.~\ref{fig3}, we demonstrate the long-time evolutions of $P_{e}(t)$ with large enough $\tau$ and different modulation parameters. Having in mind that the modulation effect disappears when $\Omega\tau=2n\pi$, here we take $\Omega\tau=(2n+1)\pi$ as an example. In Fig.~\figpanel{fig3}{a}, the atom shows evident population revivals in the absence of the modulation (i.e., $\chi=0$), but the revivals tend to fade as the modulation depth $\chi$ grows. This can be understood from the dynamical phase factor $F=\text{exp}[i\phi(t,\tau)]$, which dynamically modifies the retarded feedback term as shown in Eq.~(\ref{eq7}). The evolutions of the real part of $F$ with different values of $\chi$ are plotted in Fig.~\figpanel{fig3}{b} (the imaginary part of $F$ shows similar evolutions, which are not demonstrated here). For small $\chi$, the dynamical phase factor $F$ changes slowly within a small range that is away from zero. For large $\chi$ (with $\Omega$ remaining invariant), however, the retarded feedback term (with the prefactor $F$) oscillates rapidly with a nearly vanishing average contribution, such that the model behaves like a small atom with negligible population revivals. In view of this, the result in Fig.~\figpanel{fig3}{a} shows the possibility of protecting quantum emitters from unwanted environment backactions and controlling the non-Markovianity of the quantum dynamics~\cite{FCretard2,quantifier}.  

Moreover, we would like to interpret this result using the Jacobi-Anger expansion of the dynamical phase factor, i.e., 
\begin{equation}
\begin{split}
F&=\text{exp}(i\phi_{0})\text{exp}[2i\chi\sin{(\Omega t})]\\
&=\text{exp}(i\phi_{0})\sum_{q=-\infty}^{+\infty}J_{q}(2\chi)e^{iq\Omega t},
\end{split}
\label{eq12}
\end{equation}    
where $\Omega\tau=(2n+1)\pi$ and $\theta=0$ have been assumed and $J_{q}(\chi)$ is the Bessel function of the first kind. Equation~(\ref{eq12}) shows that the influence of the dynamical phase factor becomes negligible if $\chi$ is large enough because all $J_{q}(\chi)$ are small for $\chi\rightarrow+\infty$. This is also why in Fig.~\figpanel{fig2}{c} the modulation effect exhibits a non-monotonic behavior as $\chi$ increases. For other cases of $\Omega\tau\neq(2n+1)\pi$ and $\theta\neq0$, the expansion of $F$ becomes more complicated but the asymptotic behavior is similar.    

We also plot in Fig.~\figpanel{fig3}{c} the evolutions of $P_{e}(t)$ in the deep non-Markovian regime with two very different values of $\Omega$, i.e., $\Omega/\Gamma=5.5\pi$ and $80.3\pi$. The two evolution curves show good agreement, illustrating that the population revivals cannot be eradicated by using larger modulation frequency (here we fix the value of $\chi$ by changing $\alpha$ and $\Omega$ simultaneously, otherwise $\chi$ should decrease as $\Omega$ increases, leading to negligible modulation effects for large enough $\Omega$). This can be understood again from the evolutions of $\text{Re}(F)$ shown in Fig.~\figpanel{fig3}{d}: the dynamical phase factor oscillates faster with higher modulation frequency, yet its average contribution is almost unchanged. Although we have used $\Omega\tau=(2n+1)\pi$ in Figs.~\figpanel{fig3}{c} and \figpanel{fig3}{d}, the conclusion here also holds for other values of $\Omega\tau\neq2n\pi$. This can be seen from the inset in Fig.~\figpanel{fig3}{c}, where $P_{e}(t=11/\Gamma)$ changes slightly with $\Omega$ in a periodic manner.

Before moving to the next section, we would like to briefly discuss the influence of the additional phase difference $\varphi$ on the spontaneous emission dynamics of the atom. It can be seen from Eq.~(\ref{eq7}) that the feedback term is modified by $\varphi$ in terms of a cosine function: the amplitude of the feedback term is proportional to $\Gamma\cos{\varphi}$ in this case. In other words, such a phase difference alters the effective decay rate of the giant atom rather than introducing any new physics to the decay dynamics. When $\text{mod}(\varphi,2\pi)\neq0$, the results above can be recovered by tuning other parameters such as the atom-waveguide coupling strengths (the amplitude of the retarded feedback term can differ from that of the instantaneous decay term if the coupling strengths at the two coupling points are different~\cite{DLprr2}). However, as will be seen in the next section, such a phase difference enables an effective chiral interaction between the atom and the waveguide field and thereby leads to chiral output fields.  

\section{Chiral and tunable output fields}\label{sec4}

In experiments, the spontaneous emission of the (giant) atom can be examined by measuring the output fields at the ports of the waveguide. In view of this, it is convenient to transform the field amplitude $c_{k}(t)$ to real space via
\begin{equation}
c(x,t)=\frac{1}{\sqrt{2\pi}}\int dk c_{k}(t)e^{ikx},
\label{eq13}
\end{equation} 
whose square modulus can be measured by a photon detector placed at position $x$. To derive the real-space field amplitude in Eq.~(\ref{eq13}), we rewrite Eqs.~(\ref{eq3}) and (\ref{eq4}) as 
\begin{eqnarray}
\dot{c}_{e}&=&-i\omega(t)c_{e}-i\int_{-\infty}^{+\infty}dk g\left(1+e^{i\varphi}e^{ikd}\right)c_{k},\label{eq14}\\
\dot{c}_{k}&=&-i\omega_{k}c_{k}-ig\left(1+e^{-i\varphi}e^{-ikd}\right)c_{e}.\label{eq15}
\end{eqnarray} 
Substituting the formal solution of $c_{k}(t)$, i.e., 
\begin{equation}
c_{k}(t)=-i\int_{0}^{t}dt' g\left(1+e^{-i\varphi}e^{-ikd}\right)c_{e}(t')e^{-i\omega_{k}(t-t')},
\label{eq16}
\end{equation}
into Eq.~(\ref{eq13}), one has
\begin{widetext}
\begin{equation}
\begin{split}
c(x,t)&=\frac{-ig}{\sqrt{2\pi}}\int_{0}^{t}dt'\int dk\left(1+e^{-i\varphi}e^{-ikd}\right)e^{ikx}c_{e}(t')e^{-i\omega_{k}(t-t')}\\
&=\frac{-i\sqrt{2\pi}g}{v_{g}}\int_{0}^{t}dt' \Big[e^{ik_{0}x}\delta\left(t-t'-x/v_{g}\right)+e^{-ik_{0}x}\delta\left(t-t'+x/v_{g}\right)+e^{-i\varphi}e^{ik_{0}(x-d)}\delta\left(t-t'-x/v_{g}+d/v_{g}\right)\\
&\quad\,+e^{-i\varphi}e^{-ik_{0}(x-d)}\delta\left(t-t'+x/v_{g}-d/v_{g}\right)\Big]c_{e}(t')e^{-i\omega_{0}(t-t')}\\
&=\frac{-i\sqrt{2\pi}g}{v_{g}}\Big[e^{ix(k_{0}-\omega_{0}/v_{g})}c_{e}\left(t-x/v_{g}\right)\Theta(x)\Theta\left(t-x/v_{g}\right)+e^{-ix(k_{0}-\omega_{0}/v_{g})}c_{e}\left(t+x/v_{g}\right)\Theta(-x)\Theta\left(t+x/v_{g}\right)\\
&\quad\,+e^{-i\varphi}e^{i(x-d)(k_{0}-\omega_{0}/v_{g})}c_{e}\left(t-x/v_{g}+d/v_{g}\right)\Theta(x-d)\Theta\left(t-x/v_{g}+d/v_{g}\right)\\
&\quad\,+e^{-i\varphi}e^{-i(x-d)(k_{0}-\omega_{0}/v_{g})}c_{e}\left(t+x/v_{g}-d/v_{g}\right)\Theta(d-x)\Theta\left(t+x/v_{g}-d/v_{g}\right)\Big],
\end{split}
\label{eq17}
\end{equation}
\end{widetext}
where $\omega_{k}=\omega_{0}+(k-k_{0})v_{g}$ has been used again. If the photon detector is located at the right side of the giant atom, i.e., $x=d+l$ ($l>0$), we have
\begin{equation}
\begin{split}
|c_{R}(t)|&=|c(x=d+l,t)|=\frac{\sqrt{2\pi}g}{v_{g}}\Big|\Big[c_{e}(\tilde{t})\Theta(\tilde{t})\\
&\quad\,+e^{i(\phi_{0}'+\varphi)}c_{e}(\tilde{t}-\tau)\Theta(\tilde{t}-\tau)\Big]\Big|.
\end{split}
\label{eq18}
\end{equation} 
On the other hand, if the detector is located at the left side of the atom, i.e., $x=-l$, we have
\begin{equation}
\begin{split}
|c_{L}(t)|&=|c(x=-l,t)|=\frac{\sqrt{2\pi}g}{v_{g}}\Big|\Big[c_{e}(\tilde{t})\Theta(\tilde{t})\\
&\quad\,+e^{i(\phi_{0}'-\varphi)}c_{e}(\tilde{t}-\tau)\Theta(\tilde{t}-\tau)\Big]\Big|.
\end{split}
\label{eq19}
\end{equation} 
In Eqs.~(\ref{eq18}) and (\ref{eq19}), we have assumed $\tilde{t}=t-l/v_{g}$ and $\phi_{0}'\equiv k_{0}d-\omega_{0}d/v_{g}=\phi_{0}-\omega_{0}\tau$. Note that $v_{g}=(\partial\omega_{k}/\partial k)|_{k=k_{0}}\neq\omega_{0}/k_{0}$ and thereby $\phi_{0}\neq\omega_{0}\tau$ if $\omega_{k}$ is not exactly proportional to $|k|$~\cite{JTShen2009}. However, we have checked that the main results in this section also hold even if $\phi_{0}=\omega_{0}\tau$ (i.e., $\omega_{k}=|k|v_{g}$~\cite{linear1,linear2,linear3}). According to Eqs.~(\ref{eq18}) and (\ref{eq19}), the output fields should be asymmetric (i.e., the spontaneous emission of the atom should be chiral) if the additional phase difference $\varphi$ is not an integer multiple of $\pi$. 

\begin{figure}[ptb]
\centering
\includegraphics[width=8.5 cm]{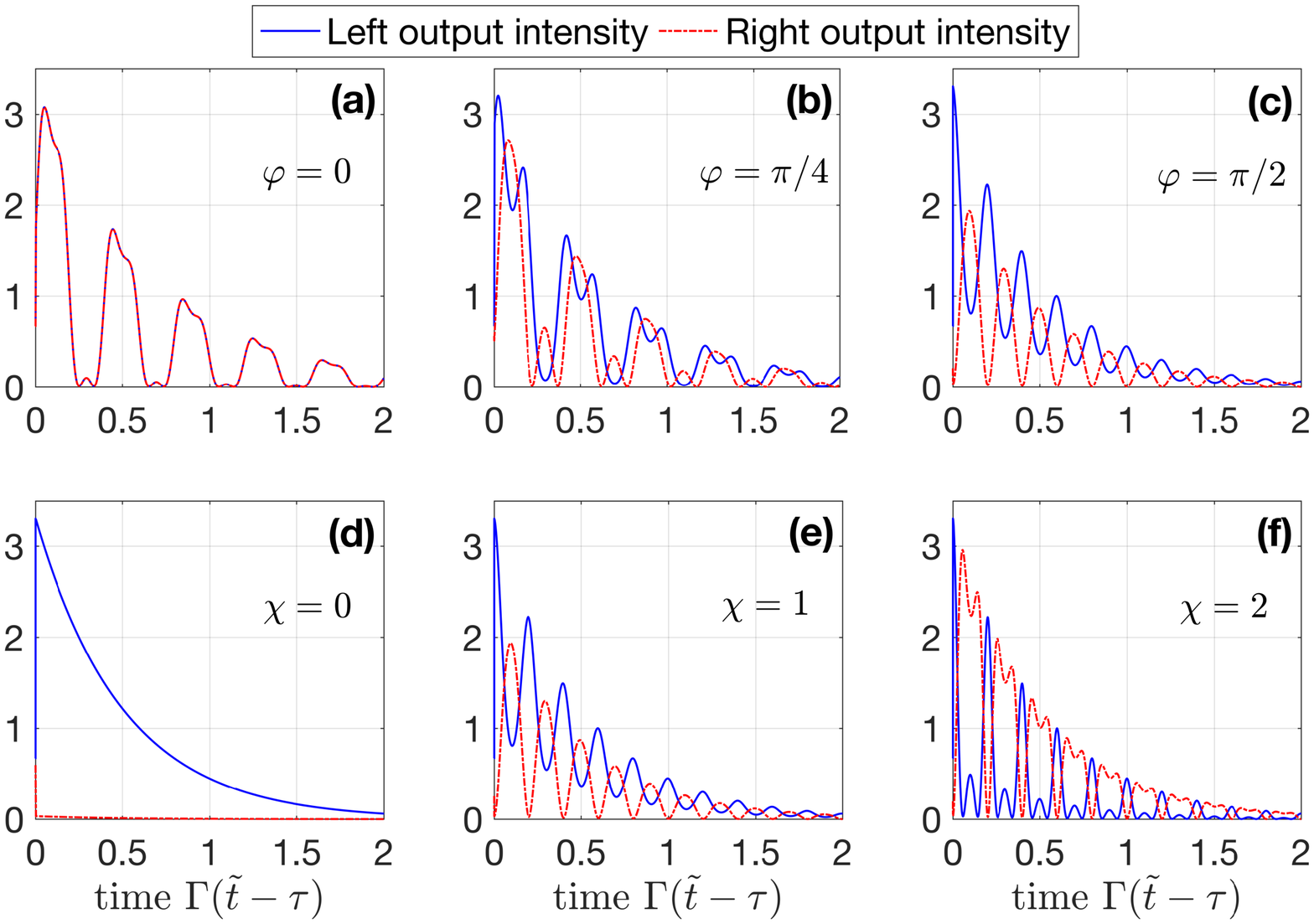}
\caption{Dynamic evolutions of the left and right output intensities (in units of $\Gamma/2v_{g}$) versus $\tilde{t}-\tau$ with different values of modulation parameters (cosine-type modulation). We assume $\chi=1$ in panels (a)-(c) and $\varphi=\pi/2$ in panels (d)-(f). Other parameters are $\phi_{0}=(2m+1)\pi$, $\phi_{0}'=(2n+1/2)\pi$, $\tau\Gamma=0.2$, $\Omega/\Gamma=5\pi$, and $\theta=0$.}\label{fig4}
\end{figure}

We first examine the evolutions of the left and right output intensities $|c_{L}(t)|^{2}$ and $|c_{R}(t)|^{2}$ versus the renormalized time $\tilde{t}-\tau$ with different values of $\varphi$ and $\chi$ (for $\tilde{t}<0$, the emitted photon cannot be detected, while for $0<\tilde{t}<\tau$, the output fields are always symmetric and exhibit no modulation effect since the retarded feedback has not come into effect). As shown in Figs.~\figpanel{fig4}{a}-\figpanel{fig4}{c}, the output fields are symmetric (achiral) if $\varphi=0$, otherwise the output fields become chiral with the chiral effect being more evident if $\text{mod}(\varphi,2\pi)=\text{mod}(\phi_{0}',2\pi)$. The output fields exhibit oscillating temporal profiles due to the cosine-type modulation~\cite{FCretard1}. The case without modulation is demonstrated in Fig.~\figpanel{fig4}{d}, where the right output is nearly inhibited and the left one shows an exponentially damped profile. Moreover, as shown in Figs.~\figpanel{fig4}{d}-\figpanel{fig4}{f}, the modulation depth $\chi$ plays an important role for harnessing the profiles of the output fields without affecting the chirality. We point out that changing the modulation phase $\theta$ leads to a slight shift of the profiles along the time axis (not shown here), which provides an additional tunability of the output fields.    

\begin{figure}[ptb]
\centering
\includegraphics[width=8.5 cm]{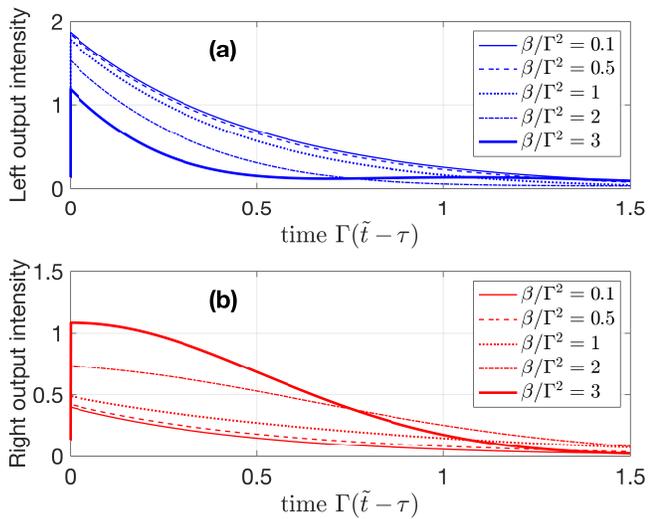}
\caption{Dynamic evolutions of the left (a) and right (b) output intensities (in units of $\Gamma/2v_{g}$) versus $\tilde{t}-\tau$ with different values of $\beta$ (linear modulation). Other parameters are $\phi_{0}=(2m+1)\pi$, $\phi_{0}'=(2n+1/2)\pi$, $\varphi=\pi/2$, and $\tau\Gamma=1$.}\label{fig5}
\end{figure} 

Besides the cosine-type modulation discussed above, one can also consider aperiodic modulations to engineer richer chiral output profiles. For example, we now consider a linear frequency modulation in the form of $\omega(t)=\omega_{0}+\beta t$, with $\beta$ being the modulation rate and having the dimension of $\text{Hz}^{2}$. Although $\omega(t)$ is not a bounded function in this case, we limit ourselves to the case of small $\beta$ and short evolution time to ensure $\beta t\ll\omega_{0}$. Now $\phi(t,\tau)$ can be written as
\begin{equation}
\phi(t,\tau)=\phi_{0}+\beta\tau\left(t-\frac{\tau}{2}\right).
\label{eq20}
\end{equation}
Equation~(\ref{eq20}) shows that the dynamical phase depends linearly on $t$, with the prefactor determined by both $\beta$ and $\tau$. Therefore we consider in this case a larger $\tau$ to achieve stronger modulation effects. 

Figure~\ref{fig5} depicts the dynamic evolutions of the left and right output intensities versus $\tilde{t}-\tau$ with different values of $\beta$. It shows that the chiral temporal profiles of the output fields can be markedly affected by the modulation rate $\beta$. In particular, the profile of the right output field can be tuned from an exponential-like shape to a Gaussian-like shape as $\beta$ changes. Moreover, the overall intensity of the left (right) output field reduces (grows) gradually with the increase of $\beta$. As a result, aperiodic frequency modulations provide richer schemes for tuning the chirality of the output dynamics.  

Finally, we would like to point out that the results in this section, which arise from the effective chiral atom-waveguide interaction and appropriate frequency modulations, are in principle compatible with other chiral quantum optical mechanisms such as spin-momentum locking~\cite{Lodahlchiral,Zollerchiral}, topological reservoirs~\cite{VegaPRA,topobath}, and synthetic gauge fields~\cite{sawtooth}. The chirality of our proposal is tunable \emph{in situ}~\cite{WXchiral2}, which makes it an excellent alternative of chiral quantum interfaces~\cite{AFKchiral}. In analogy to the giant-atom structures, chiral quantum interfaces have also been created by engineering composite interactions (containing, e.g., linear and nonlinear interactions) for small-atom dimers, without breaking the Lorentz reciprocity of the system~\cite{npjZoller}. Such systems, however, might call for simultaneous frequency modulations for both atoms.

\section{Conclusions} 

In summary, we have studied the spontaneous emission dynamics of a two-level giant atom with modulated transition frequency. We have revealed that the non-Markovian retardation effect, which stems from the nonnegligible time delay of photons traveling between different coupling points, endows the giant-atom interference effect with a dynamical modification. This thus allows for controlling the spontaneous emission of the atom, depending on both the concrete form of the frequency modulation and the value of the time delay. As an example, we have considered a cosine-type frequency modulation and studied in detail its influence on the dynamic evolutions of the atomic population. Based on the controllable spontaneous emission, we have also demonstrated how to engineer chiral output fields with tunable temporal profiles. This can be achieved by introducing an additional phase difference between the two atom-waveguide coupling coefficients and using various modulation schemes. 

The scenario in this paper can be immediately extended to situations of a single multilevel giant atom~\cite{LambAFK,DLprr,DLlambda}, multiple correlated two-level giant atoms~\cite{NoriGA,braided,DLretard,LvGA}, and structured baths~\cite{ZhaoWbound,VegaPRA,topobath,AFKstructure,ZhangSfop}. For the latter situation, it is also possible to realize efficient dipole-dipole interactions resorting to appropriate frequency modulations even if the atoms are detuned from each other~\cite{arxivClerk}. Furthermore, the results in this paper can be extended beyond the single-excitation space by, e.g., considering extra coherent pumping for the atom, which may allow for more exotic quantum phenomena~\cite{beyond,clock}. Potential applications of our proposal include but are not limited to: (i) providing a new wisdom for controlling the non-Markovianity of quantum dynamics (in terms of the controlled atomic population revivals) and improving the performance of the preexisting non-Markovianity quantifiers~\cite{FCretard2,quantifier}; (ii) creating chiral single-photon pulses with highly tunable temporal profiles; (iii) engineering more advanced quantum switches which are based on the interference between the atomic spontaneous emission and the propagating field in the waveguide; (iv) developing quantum simulation techniques based on giant atoms, such as simulations of open (chiral) many-body systems~\cite{braided,AFKchiral}.  

\section*{Acknowledgments} 

This work was supported by the National Natural Science Foundation of China (under Grants No. 12074030 and No. 12274107) and the Science Foundation of the Education Department of Jilin Province (under Grant No. JJKH20211279KJ).

\end{document}